\documentclass[usenatbib,useAMS]{mn2e}

\def\gapprox{\lower.4ex\hbox{$\;\buildrel >\over{\scriptstyle\sim}\;$}}
\def\lapprox{\lower.4ex\hbox{$\;\buildrel <\over{\scriptstyle\sim}\;$}}

\def\bk{\mbox{\boldmath $k$}}

\def\bx{\mbox{\boldmath $x$}}

\def\bu{\mbox{\boldmath $u$}}
\def\bv{\mbox{\boldmath $v$}}

\def\bp{\mbox{\boldmath $p$}}
\def\bB{\mbox{\boldmath $B$}}
\def\bE{\mbox{\boldmath $E$}}

\title[Magnetic field amplification at SN shocks]{Saturated magnetic field amplification at supernova shocks}

\author[Q. Luo, D. Melrose]
      {Qinghuan Luo and Don Melrose\\
         School of Physics, The University of Sydney, NSW 2006, Australia}
\date{
          --- Received in original form February, 2009
        }

\pubyear{2009}
\begin{document}

\maketitle

\begin{abstract}
Cosmic-ray streaming instabilities at supernova shocks are discussed in the quasilinear 
diffusion formalism which takes into account the feedback effect of wave growth on 
the cosmic ray streaming motion. In particular, the nonresonant instability that leads to 
magnetic field amplification in the short wavelength regime is considered. The linear growth 
rate is calculated using kinetic theory for a streaming distribution. We show that
the nonresonant instability is actually driven by a compensating current in the background 
plasma. The nonresonant instability can develop into a nonlinear regime
generating turbulence. The saturation of the amplified magnetic fields 
due to particle diffusion in the turbulence is derived analytically. It is shown that
the evolution of parallel and perpendicular cosmic-ray 
pressures is predominantly determined by nonresonant diffusion. However, 
the saturation is determined by resonant diffusion which tends to reduce the 
streaming motion through pitch angle scattering. The saturated level can exceed the mean 
background magnetic field.
\end{abstract}

\begin{keywords}
ISM: magnetic field -- cosmic rays -- supernova remnants -- 
shock -- radiation mechanisms
\end{keywords}

\section{Introduction}

Diffusive shock acceleration (DSA) is regarded as the preferred mechanism for the 
acceleration of the Galactic cosmic rays (CRs) \citep{d83}. It is commonly believed
that SN shocks are responsible for acceleration of high energy CRs 
at least up to the `knee' ($\sim4\times10^{15}\,\rm eV$)
of the CR spectrum \citep{h06}. In the DSA model, particles gain energy 
by bouncing back and forth across the shock. Although particles only gain a small amount of
energy in each crossing, they can be accelerated to very high energy through
many crossings provided they can be trapped in the acceleration region for a sufficiently 
long time. The acceleration is rather efficient and naturally leads to a power-law energy
distribution. Despite these advantages, there have been two long standing problems. 
First, the standard DSA theory predicts the maximum energy
well below the `knee' \citep{b04}. The maximum energy achievable is limited by
both the Bohm approximation, in which the particle's gyroradius must not exceed the mean 
free path to scattering, and the condition that the gyroradius is smaller than the
shock's width (otherwise the particle would escape from the acceleration region)
\citep{b04,zetal08}.  For a magnetic field of order $10^{-10}\,\rm T$ 
and typical parameters of SN shocks, one has the maximum energy $\sim 5\times 10^{14}\,\rm eV$.
Second, turbulence is required for effective scattering in the acceleration region
so that the particles can be trapped \citep{s75,b78,lc83a,lc83b}. So far, 
in the standard DSA theory, one generally postulates that Alfv\'en 
turbulence can be generated by CRs themselves through resonant 
interactions, even though in practice the growth of Alfv\'en waves
due to CRs is known to be ineffective. 

There is growing interest in the possibility  that the magnetic field 
at the shock may be amplified due to CR induced instability 
in the nonresonant regime \citep{mv82,bl01,b04}. \citet{b04} showed 
a nonresonant form of the instability driven by a CR current 
can outgrow the familiar resonant form, leading to magnetic amplification. 
A strong magnetic field at shocks reduces the gyroradius and this raises the 
maximum energy to which particles can be accelerated. 
X-ray observations of young SNRs suggest that the magnetic field strength 
near the shocks is much higher than that in the ISM
and is a strong function of the shock speed \citep{vl03,vetal08}. 
The presence of strong magnetic fields may explain
the lack of strong TeV gamma-ray fluxes, as shown from HESS observations \citep{vetal08}.
The nonthermal X-rays are well described by synchrotron spectra.
A stronger magnetic field in the emission region implies that a lower electron 
energy is required. This would lead to a lower TeV gamma flux from inverse Compton scattering. 

The nonresonant growth of Alfv\'en waves due to CR streaming has
been discussed in both the MHD formalism \citep{bl01,b04,zetal08} and 
the kinetic formalism \citep{m05,retal06,ab08}.
In both formalisms, the instability is shown to exist in the linear regime.
To test the magnetic field amplification model against observations one needs
to determine the saturated magnetic field accurately. The saturation of the instability
cannot be determined in linear theory as the reaction of the instability on the 
CR momemtum distribution is not automatically included in the linear calculation. 
Although there are numerical simulations
of the instability that extend to the nonlinear regime, different saturation levels have been
predicted \citep{netal08,rs08}. In this paper, the instability is discussed in the quasilinear formalism
in which the reaction of the instability on the CR distribution can be
included self-consistently. So, in this formalism one can estimate the saturation analytically,
with both nonresonant and resonant diffusion processes considered. 
The treatment of the nonresonant diffusion presented here is similar to that 
used for the firehose instability \citep{d72}. We emphasize the
major difference between the CR streaming instability and 
the firehose instability. The former is caused by streaming motion and the latter is due to
a pressure anisotropy with excess of parallel pressure over the perpendicular 
pressure (with respect to the mean magnetic field). To some extent, the 
streaming instability also resembles the Weibel instability---a nonresonant, purely
growing mode driven by anisotropy in the particle distribution \citep{w59}. 

In Sec 2 the kinetic theory of CR streaming instabilities is discussed with emphasis
on the nonresonant instability. Quasilinear diffusion driven by the nonresonant instability 
is discussed in both the short and long wavelength approximations in the nonresonant regime
in Sec 3 and in the resonant regime in Sec 4. Application to SN shocks is discussed in Sec 5.

\section{Cosmic-ray streaming instabilities}

We outline the kinetic theory of CR streaming instabilities including both the usual resonant 
instability and the nonresonant instability and focus particularly on the latter. 
Our treatment builds on other recent discussions of linear kinetic theory of the CR-induced 
nonresonant instability \citep{retal06,ab08}. For convenience we assume a single species of 
CRs with charge $q$ and mass $m$.

\subsection{CR streaming motion}

To model CR streaming at velocity $v_{CR}$ along the mean magnetic field
we consider a class of streaming distributions in momentum space given by
\begin{equation}
f(u_\parallel,u_\perp)=n_{CR}\Biggl[1+(2\sigma+1){v_{CR}\over v}\left({u_\parallel\over u}
\right)^{2\sigma-1}\Biggr]{g(u)\over 4\pi u^2},
\label{eq:fcr}
\end{equation}
where $u_\parallel$ and $u_\perp$ are the nondimensional momenta (normalized by $mc$) 
parallel and perpendicular to the mean magnetic field, respectively, 
$\sigma$ is an integer $\geq1$, $n_{CR}$ is the CR number density,
$v$ is the CR's velocity written as a function of $u_\parallel$ and 
$u_\perp$, $u=(u^2_\perp+u^2_\parallel)^{1/2}$, and
\begin{equation}
g(u)=\left\{
\begin{array}{lr}
\displaystyle{
{p-1\over u_1}\left[
1-\left({u_1\over u_2}\right)^{p-1}\right]^{-1}\left({u\over u_1}\right)^{-p}},&
u_1\leq u\leq u_2,\\
0&{\rm otherwise.}
\end{array}
\right.
\label{eq:gp}
\end{equation}
For the standard DSA one has $p=2$ \citep{b78,d83} and when the nonlinear effect on DSA is
included, $p$ deviates from this canonical value \citep{e84}.
The distribution with $\sigma=1$ corresponds to that used in
\citet{m86}. The distribution (\ref{eq:fcr}) implies 
\begin{eqnarray}
n_{CR}&=&4\pi\int^\infty_0u_\perp du_\perp\int^\infty_{-\infty}
du_\parallel\, f(u_\parallel,u_\perp)\nonumber\\
&=&
2\pi\int^\pi_0\sin\alpha d\alpha\int^\infty_0
u^2du\, f(u\cos\alpha,u\sin\alpha),
\label{eq:ncr}
\end{eqnarray}
where $\cos\alpha\equiv u_\parallel/u$. In the second expression in (\ref{eq:ncr}), 
one chooses $(u,\alpha)$ in place of $(u_\parallel,u_\perp)$ as independent variables. 
It can be verified that averaging the parallel velocity $v_\parallel$ over the 
distribution (\ref{eq:fcr}) gives the streaming velocity $v_{CR}$,
which is independent of the choice of the parameter $\sigma$.
The CR current is then given by $J_{CR}=qn_{CR}v_{_{\rm CR}}$. The presence of streaming 
CRs affects the background plasma in two ways, due to their charge density and their
current density, respectively. The background plasma must have a charge density and a 
current density that are equal and opposite to those of the CRs. This requires that the 
electrons and ions (assumed to be protons) have different charge densities, $n_e\ne n_p$, 
and that they move relative to each other with streaming velocities $v_e\ne v_p$,
which are assumed to be along the guiding magnetic field. The neutralization conditions require
 \citep{a83} 
\begin{equation}
e(n_e-n_p)=qn_{_{\rm CR}},\quad
 e(n_ev_e-n_pv_p)=qn_{_{\rm CR}}v_{_{\rm CR}}.
\label{eq:neutral}
\end{equation}
These properties of the background plasma drive the nonresonant instability attributed to the CRs.

\subsection{Dispersion relation}

A formal procedure to derive the dispersion relation involves separating the plasma 
response tensor into that for a background plasma, denoted by 
$K_{ij}$, plus that for the CR component, denoted by $\Delta K_{ij}$.
The background plasma can be regarded as  a cold, magnetized plasma, while the CR component
is described by the distribution (\ref{eq:fcr}). Assume that the gyrofrequency of CRs is
$\Omega=|q|B/m$, where $B$ is the mean background magnetic field. A useful approximation is
$k_\perp v_\perp/\tilde{\Omega}\ll1$, where $k_\perp$ is the perpendicular wave number,
$\tilde{\Omega}=\Omega/\gamma$ and $\gamma=(1+u^2)^{1/2}$ is the Lorentz factor of CRs. 
The approximation implies that in the response tensor, only the first gyroharmonics 
terms are important. For the background plasma, one assumes $v^2_A\ll c^2$ and 
the low-frequency approximations, $\omega\ll\Omega_i$, $\omega\ll |k_\parallel v_\parallel|$ and $\omega\ll
\tilde{\Omega}$, where $\Omega_i$ is the gyrofrequency of ions in the background plasma. 
The background magnetic field is assumed to be along the 3-axis. Since $K_{33}\propto \Omega^2_i/\omega^2$
can be set to $\infty$, only the $2\times2$ components of the response tensor
are relevant. For the background plasma, these components can be written as
\begin{eqnarray}
K_{11}&=&K_{22}={c^2\over v^2_A},\\
K_{12}&=&-K_{21}=i{c^2\over v^2_A}{\Omega_i\over\omega}
\biggl[{n_e-n_p\over2n_i}+{k_\parallel(n_ev_e-n_pv_p)\over\omega n_i}
\biggr],
\label{eq:K12a}
\end{eqnarray}
where $n_\parallel=k_\parallel c/\omega$,
$v_A=\Omega_i/\omega_{pi}$ is the Alfv\'en speed, $\omega_{pi}$
is the ion plasma frequency of the background plasma.
Using the neutralization conditions (\ref{eq:neutral}),  
(\ref{eq:K12a}) can be rewritten in the form
\begin{eqnarray}
K_{12}=-K_{21}=-i\eta n^2_\parallel \chi^{(0)}_h,\quad
\chi^{(0)}_h={\mu_0J_{CR}\over2k_\parallel B}.
\end{eqnarray}
The CR components are
\begin{eqnarray}
\Delta K_{11}=\Delta K_{22}=- in_\parallel^2\chi_a,\quad
\Delta K_{12}=-\Delta K_{21}= -in_\parallel^2\chi_h,
\end{eqnarray}
where
\begin{eqnarray}
\chi_a&=&{\pi\over2}A
\!\int^{x_1}_{x_2}
dx\,x^{p+2\sigma-3}(1-x^2)H(1-x^2),\label{eq:chia}\\
 \chi_h&=&A
\!\int^{x_1}_{x_2}
dx\,x^{p+2\sigma-3}\biggl[(1-x^2)\ln\left|{1+x\over 1-x}\right|+2x\nonumber\\
&&-4\sum^{\sigma-1}_{s=1}{x^{1-2s}\over4s^2-1}\biggr],
\label{eq:chih}
\\
A&=&{\mu_0J_{CR}\over 4k_\parallel B}
{(4\sigma^2-1)(p-1)\over x^{p-1}_1},
\end{eqnarray}
with $(x_1,x_2)=b(1/u_1,1/u_2)$, $b=\Omega/|k_\parallel|c$ and $H(x)$ the step function.

Eq (\ref{eq:chia}) and (\ref{eq:chih}) correspond to the dissipative and reactive part
respectively of the response tensor due to CRs. 
In the strong magnetic field limit $x>1$, the dissipative part is zero. 
When $x<1$, both dissipative and reactive terms contribute to the dispersion. 
The dissipative part is due to CRs in the resonance $x\equiv b/u\approx \mu$ as discussed extensively 
in the context of the resonant instability (see discussion below) \citep{kp69,mw70,s75,mv82,lc83a}. 
The dispersion relation takes the following form
\begin{eqnarray}
\omega^2&=&{\textstyle{1\over2}}
\Biggl\{{k^2\over k^2_\parallel}+1+2i\chi_a
\nonumber\\
&&\pm\biggl[{k_\perp^4\over k^4_\parallel}
+4\Bigl(\chi^{(0)}_h+\chi_h\Bigr)^2\biggr]^{1/2}
\Biggr\}k^2_\parallel v^2_A.
\label{eq:disp}
\end{eqnarray}
In the $\chi_a\to0$ and $\chi_h\to 0$ limit ($\chi^{(0)}_h$ must also be zero), 
(\ref{eq:disp}) with the lower sign reproduces the usual Alfv\'en mode 
dispersion $\omega=|k_\parallel| v_A$ and with the upper sign the fast mode 
dispersion $\omega=kv_A$.

\subsection{Streaming instabilities}

Instability occurs when $\omega$ has an imaginary part of the appropriate sign. 
An imaginary part can arise in two different ways, which we refer to as resonant and nonresonant 
instabilities. Writing $\omega\to\omega+i\Gamma$ in (\ref{eq:disp})
and assuming $\Gamma\ll\omega$, the resonant instability is described by equating 
the small imaginary terms, giving $\Gamma=\chi_ak_\parallel^2v_A^2/2\omega$. The value 
of $\chi_a$ is determined by the CR distribution. 
A resonant instability can develop for both Alfv\'en and fast-mode waves if streaming 
CRs are in resonance with the wave considered. 
For Alfv\'en waves ($\omega=|k_\parallel|v_A$), the growth rate is
\begin{equation}
\Gamma\approx {\textstyle{1\over2}}\chi_a|k_\parallel| v_A.
\end{equation}
The polarization of both modes is generally elliptical and becomes
approximately circular for nearly parallel propagation $|k_\perp/k_\parallel|\ll1$.
In each mode, waves of both sense of polarization can grow.

On neglecting $\chi_a$, (\ref{eq:disp}) with the upper sign becomes a real equation, because the 
quantity inside the square root is a sum of squares. For the
negative sign (Alfv\'en waves), $\omega^2$ can be negative, and one of the two solutions corresponds to an
intrinsically growing wave. This is identified as the nonresonant instability.
For $|k_\perp/k_\parallel|\ll1$, the condition for this nonresonant instability 
to occur is $(\chi^{(0)}_h+\chi_h)^2>1$.
The growth rate is 
\begin{equation}
\Gamma\approx\sqrt{2}\left(|\chi^{(0)}_h+\chi_h|-1\right)^{1/2}|k_\parallel|v_A.
\label{eq:Gam}
\end{equation}
In principle, this can be satisfied either due to $|\chi_h|>1$ or to $|\chi_h^{(0)}|>1$. 
For $\sigma=1$, the integral $\chi_h$ given by 
(\ref{eq:chih}) can be calculated exactly, i.e.,
\begin{eqnarray}
\chi_h&=&{3\mu_0J_{CR}(b-1)\over4k_\parallel Bx^{p-1}_1}\int^{x_1}_{x_2}dx\,
x^{p-1}\Biggl[(1-x^2)\ln\left|{1+x\over1-x}\right|+2x
\Biggr]\nonumber\\
&=&{3\mu_0J_{CR}\over16k_\parallel Bx_1}\Biggl\{2\biggl[x_1(1+x^2_1)
-x_2(1+x^2_2)\biggr]\nonumber\\
&&-(x^2_1-1)^2\ln\left|{1+x_1\over1-x_1}\right|
+(x^2_2-1)^2\ln\left|{1+x_2\over1-x_2}\right|\Biggr\}.
\label{eq:chih2}
\end{eqnarray}
The second equality in (\ref{eq:chih2}) is obtained for $p=2$.
The condition $|\chi_h|>1$ cannot be satisfied for $x_1\ll1$.
Thus, only the second possibility ($\chi^{(0)}_h>1$) is relevant.

For $x_2\ll x_1\ll 1$, one finds
\begin{equation}
\chi_h\approx {\mu_0J_{CR}x^2_1\over4k_\parallel B},
\label{eq:chih3}
\end{equation}
so that the contribution from the CR current is much smaller than 
the contribution due to the compensating current in the background 
plasma. It follows that in treating the nonresonant instability, one can neglect 
the direct contribution of the CRs, described by $\chi_h$, in comparison 
with the indirect contribution, described by $\chi_h^{(0)}$.
The growth rate (\ref{eq:Gam}) can be written 
in the following approximate form
\begin{equation}
\Gamma\approx \sqrt{2}\left(\chi^{(0)}_h-1\right)^{1/2}k_\parallel v_A=
\sqrt{2}\left({\mu_0J_{CR}\over k_\parallel B}-1\right)^{1/2}k_\parallel v_A.
\label{eq:Gam2}
\end{equation}
Nonresonant instability requires 
\begin{equation}
|J_{CR}|>{k_\parallel B\over\mu_0}.
\label{eq:condition}
\end{equation}
An explanation for why the direct contribution from the CR current is so
much smaller than the indirect contribution from the background plasma is that 
in the limit $k_\parallel r_g\gg1$, most of CRs move rather rigidly.
Their only role in this limit is to induce the compensating current in the 
background plasma that drives the instability. It is appropriate to point out 
here that a small fraction of CRs can satisfy the resonant condition ($\mu=x$)
and that resonant instability due to these resonant CRs is insignificant
compared to the nonresonant instability due to the compensating current. 
However, the resonant interactions 
between these CRs and waves with $k_\parallel\gg1/r_g$ can lead to resonant 
diffusion that has feedback effects on CR streaming motion (cf. Sec. 4). 

\section{Quasilinear diffusion}

To determine saturation we calculate the back reaction of wave growth 
on CRs in the quasilinear diffusion theory \citep{ss64,d72}. We assume that 
the nonresonant instability discussed in Sec 2 can develop well into the nonlinear regime, 
producing turbulence. The growing turbulence causes CRs to diffuse in momentum space and
this in turn reduces the anisotropy, suppressing the instability. 

We calculate the particle diffusion in the weak 
turbulence approximation. 
The distribution of CRs in momentum space can be 
written as $f(\bu,t)=F(u_\parallel,u_\perp,t)+f^{(1)}(\bu,t)$,
where $f^{(1)}$ is a linear function of the electric and magnetic fields of the wave
and $F$ is the mean distribution, which is assumed to be axisymmetric with respect to the mean
magnetic field, $\bB$, and as the zeroth order approximation it can be taken to be 
(\ref{eq:fcr}). The response of particles to wave growth is determined by including
quadratic terms of the electric and magnetic fields of waves.
The diffusion equation for $F$ can be obtained from the Vlasov equation
(the derivation is outlined in Appendix). We assume $k_\perp r_g\ll1$ so that 
only the first gyroharmonics terms are relevant. This condition is generally satisfied for 
waves propagating nearly parallel to the mean background magnetic field.

\subsection{Nonresonant diffusion equation}

For a nonresonant instability, it is relevant to consider the approximation that the resonant width
is much larger than the growth rate, i.e., $|k_\parallel v_\parallel-\Omega/\gamma|\gg\Gamma$, 
where $\Gamma>0$ is the growth rate and for convenience the frequency is set to 
zero. The diffusion equation can be expanded
in $\Gamma/|k_\parallel v_\parallel-\Omega/\gamma|$. Let $\delta B_k$ be
the spatially Fourier-transformed magnetic fluctuations arising from the instability.
The energy of magnetic fluctuations in an elementary phase volume $d\bk/(2\pi)^3$ is
$|\delta B_k|^2/2\mu_0$. The expansion yields the following approximate diffusion equation 
(Appendix A) 
\begin{eqnarray}
{\partial F\over \partial t}&\approx&
{1\over2V}\int{d\bk\over(2\pi)^3}
\Gamma{|\delta B_k|^2\over B^2_*}
\biggl[u^2_\perp{\partial\over\partial u_\parallel}\Phi_2
{\partial\over\partial u_\parallel}
\nonumber\\
&&
-\bigl(\Phi_1+\Phi_2\bigr){u_\parallel\over u_\perp}
{\partial\over\partial u_\perp}u^2_\perp{\partial\over\partial u_\parallel}\nonumber\\
&&- u_\perp{\partial\over\partial u_\parallel}\bigl(\Phi_2-\Phi_1\bigr)
u_\parallel{\partial\over\partial u_\perp}
\nonumber\\
&&+\Phi_2{u^2_\parallel\over u_\perp}{\partial\over\partial u_\perp}
u_\perp{\partial\over\partial u_\perp}
\biggr]F,
\label{eq:dEq1}
\end{eqnarray}
\begin{eqnarray}
\Phi_1={1\over u_\parallel^2-b^2},\quad \Phi_2={u^2_\parallel+b^2\over (u^2_\parallel-b^2)^2},
\end{eqnarray}
with $b=B/B_*$, $B_*=|k_\parallel| mc/|q|$ and $V$ the volume of the region considered.

\subsection{Perpendicular and parallel kinetic energy}

It is of interest to examine the evolution of the parallel and perpendicular kinetic
energy by evaluating the rate of change in the average $u^2_\parallel$ and $u^2_\perp$ (over the mean
distribution $F$). The averaging involves singular terms $1/(u_\parallel\pm b)^n$ with $n=1,2,3$.
The $n\geq2$ terms can be avoided using
\begin{eqnarray}
\Phi_2&=&-\Phi_1+{u^2_\parallel\over b}{\partial\Phi_1\over\partial b},
\\
{\partial \Phi_1\over\partial u_\parallel}&=&
-{u_\parallel\over b}{\partial \Phi_1\over\partial b},\\
{\partial \Phi_2\over\partial u_\parallel}&=&
{3u_\parallel\over b}{\partial \Phi_1\over\partial b}-{u^3_\parallel\over b}
{\partial\over\partial b}{1\over b}
{\partial \Phi_1\over\partial b}.
\end{eqnarray}
Multiplying (\ref{eq:dEq1}) by $u^2_\parallel$ and $u^2_\perp$, respectively, and integrating
them over $2\pi u_\perp du_\perp du_\parallel$ divided by $n_{CR}$, one obtains
\begin{eqnarray}
{d\langle u^2_\parallel\rangle\over dt}&=&{1\over2\tau}
\Biggl[-\left\langle(u^2_\perp-4u^2_\parallel)\Phi_1\right\rangle
+{2\over b}{\partial\over \partial b}\left\langle(2u^2_\perp-u^2_\parallel)u^2_\parallel\Phi_1\right\rangle\nonumber\\
&&- {1\over b}{\partial\over\partial b}{1\over b}{\partial\over\partial b}\left\langle
u^4_\parallel\Phi_1u^2_\perp\right\rangle\Biggr],\label{eq:dpp}\\
{d\langle u^2_\perp\rangle\over dt}&=&
{1\over2\tau}\Biggl[-2\left\langle\Phi_1u^2_\parallel\right\rangle
+{1\over b}{\partial\over\partial b}\left\langle u^2_\parallel\Phi_1(2u^2_\parallel-
3u^2_\perp)\right\rangle
\nonumber\\
&&
+{1\over b}{\partial\over\partial b}{1\over b}{\partial\over\partial b}\left\langle
u^4_\parallel\Phi_1u^2_\perp\right\rangle\Biggr],
\label{eq:dpep}
\end{eqnarray}
where
\begin{equation}
{\tau}=
\left({1\over V}\int{d\bk\over(2\pi)^3}\,2\Gamma{|\delta B_k|^2\over
B^2_*}\right)^{-1}.
\label{eq:tau}
\end{equation}
The ratio of the parallel to perpendicular pressures is 
determined by the same angular averages as the ratio of 
$\langle u^2_\parallel\rangle$ to $\langle u^2_\perp/2\rangle$. 
It follows that  the ratio of Eqs (\ref{eq:dpp}) and (\ref{eq:dpep})
effectively determines how the ratio of the parallel to perpendicular pressures
evolves.

\subsection{Long wavelength approximation}

In the long wavelength approximation $1/k_\parallel\gg r_g$,
one has $b\gg|u_\parallel|$ and
\begin{equation}
\Phi_1\approx-\Phi_2\approx -{1\over b^2}.
\end{equation}
The rates of change in the average of $u^2_\parallel$ and $u^2_\perp$
is given by
\begin{eqnarray}
{d\langle u^2_\parallel\rangle\over dt}&=&
-\Bigl(2\langle u^2_\parallel\rangle-
\langle u^2_\perp\rangle\Bigr){1\over\tau'},\label{eq:dpp1}\\
{d\langle u^2_\perp\rangle\over dt}&=&
\langle u^2_\parallel\rangle {1\over\tau'},
\label{eq:dpep1}
\end{eqnarray}
where $\tau'$ has the same form as (\ref{eq:tau}) but with $B_*$ replaced by $B$.
Thus, one has $1/\tau'=\partial(\delta B^2/B^2)/\partial t$. 
One may compare the nonresonant diffusion to the firehose
instability. Similar to the firehose instability, a growing wave causes
the parallel kinetic energy to decrease and the perpendicular energy
to increase with time. For the CR streaming distribution (\ref{eq:fcr}),
one can show that the right-hand sides of (\ref{eq:dpp1}) and (\ref{eq:dpep1}) are zero. In the 
long wavelength regime an instability can arise from both dissipative (resonant) and reactive 
(nonresonant) effects. As we focus on the nonresonant instability in the short 
wavelength regime, quasilinear diffusion in the long wavelength regime is not 
discussed further.

\subsection{Short wavelength approximation}

The right-hand side of Eq (\ref{eq:dpp}) and (\ref{eq:dpep}) can be evaluated
in a general case in which the particle distribution is isotropic in pitch angles.
The angular integration in (\ref{eq:dpp}) and (\ref{eq:dpep}) leads to
\begin{eqnarray}
{d\langle u^2_\parallel\rangle\over dt}&=&
{1\over\tau}\Biggl\langle2+x\ln\left|{1-x\over1+x}\right|\Biggr\rangle,\label{eq:dpp2}\\
{d\langle u^2_\perp\rangle\over dt}&=&
-{1\over\tau}\Biggl\langle{2\over3}+2x^2+x^3\ln\left|{1-x\over1+x}\right|\Biggr\rangle,
\label{eq:dpep2}
\end{eqnarray}
where one uses the following integral
\begin{eqnarray}
I_{2n}&\equiv&
u^2\int^1_{-1}\mu^{2n}\Phi_1d\mu\nonumber\\
&=&2\sum^n_{s=1}
{x^{2(n-s)}\over 2s-1}+x^{2n-1}\ln\left|{1-x\over1+x}\right|.
\label{eq:I}
\end{eqnarray}
The second expression in (\ref{eq:I}) is obtained by retaining the Cauchy principal value.
Surprisingly, Eq (\ref{eq:dpp2}) and (\ref{eq:dpep2}) have
an opposite sign compared to their counterpart in the long wavelength
regime (Eq \ref{eq:dpp1} and \ref{eq:dpep1}). Although both the firehose 
instability and the instability discussed here
are driven by anisotropy in the particle distribution in momentum space,
there is an important difference between the two. 
The firehose instability is driven by anisotropy in  kinetic energy  with 
$\langle u^2_\parallel\rangle>\langle u^2_\perp\rangle/2$. By constrast, 
the streaming instability is due to the CR streaming motion ($\langle u_\parallel\rangle\neq0$),
which is also called streaming anisotropy,  and the instability
can develop even when the CR pressure is isotropic. With the specific choice of the distribution 
(\ref{eq:fcr}), one can show that $\langle u^2_\parallel\rangle=\langle
u^2_\perp\rangle/2$. Since the current in (\ref{eq:fcr}) is $\propto \mu^{2\sigma-1}$, 
it does not contribute to the average $\langle Y\rangle$ if $Y$ is proportional to an
even power of $\mu$. 

The instability can lead to pressure anisotropy with perpendicular pressure increasing.   
In the approximation $x\ll1$, one has $d\langle u^2_\parallel\rangle/dt\approx2/\tau$ and
$d\langle u^2_\perp\rangle/dt\approx-2/3\tau$.
This gives 
\begin{equation}
{d\over dt}(2\langle u^2_\parallel\rangle-\langle u^2_\perp\rangle)={\textstyle{10\over3}}\tau^{-1}>0.
\end{equation}
One should emphasize here that development of such anisotropy in kinetic energy 
is driven by the streaming motion of CRs.

From (\ref{eq:dEq1}), one may calculate the rate of change in the streaming speed due to 
nonresonant diffusion. One starts with a particle distribution with streaming motion similar to
(\ref{eq:fcr}) but without specifying the specific form of $g(p)$. Consider
the case $\sigma=1$, i.e., the streaming component is $\propto \mu$.
As in Sec 4.4, in the averaging, one can carry out the angular integration first.
This gives
\begin{eqnarray}
{d\langle v_\parallel\rangle\over dt}
&\approx& -{1\over4\tau}
\biggl\langle {v\over u^2}\biggl\{-(4+\beta^2)I_2-(2-8\beta^2+3\beta^4)I_4\nonumber\\
&&-3\beta^2(1-\beta^2)I_6
+{1\over x}{\partial\over\partial x}\Bigl[-3I_2+(1+5\beta^2)I_4
\nonumber\\
&&+\beta^2(6-7\beta^2\beta^2)I_6+3\beta^4I_8
\Bigr]\nonumber\\
&&+{1\over x}{\partial\over \partial x}{1\over x}
{\partial \over\partial x}\Bigl[I_4-\beta^2I_6-(1-\beta^2)I_8)
\Bigr]
\biggr\}\biggr\rangle,
\label{eq:vcr1}
\end{eqnarray}
where $I_{2n}$ is given by (\ref{eq:I}).
Since $dv_{_{\rm CR}}/dt=d\langle v_\parallel\rangle/dt$, 
in the limit $x\ll1$, one has
\begin{equation}
{dv_{_{\rm CR}}\over dt}\approx-{3v_{_{CR}}\over\tau}{2\over5}\biggl\langle
{1\over u^2}\Bigl(-70+28\beta^2+3\beta^4\Bigr)
\biggr\rangle.
\label{eq:vcr1b}
\end{equation}
One obtains $d v_{_{\rm CR}}/dt>0$, contrary to what one would expect. In Sec 4, cf.
Eq (\ref{eq:ratio}), we 
show that that resonant diffusion has dominant effects on the evolution 
of CR streaming motion, which ensures $d v_{_{\rm CR}}/dt<0$.

\section{Resonant diffusion}

Waves generated from the nonresonant CR-streaming instability can interact with 
CRs in resonance at $\mu=x$. Although the fraction of CRs in the resonance is small,
we show that diffusion through such resonant interactions 
has a dominant effect on the CR streaming motion. Provided that the phase 
speed of the wave is $\ll c\mu$, resonant diffusion is equivalent to 
pitch angle scattering, in which particles change their direction of motion without 
gaining or losing energy. Since the nonresonant instability is in the regime
$x\ll1$, resonant scattering involves mainly CRs with pitch angles  
$\alpha\sim\pi/2\pm x$. Since one deals with purely growing waves that can be regarded
as a quasi-mode with $\omega\to 0$, the assumption of low phase speed $\ll c\mu$ is valid 
even for particles with pitch angles near $\pi/2$. The standard formalism involves writing down the
diffusion equation involving pitch angles only \citep{m86}.
Using $F(u,\mu)=F(u_\parallel,u_\perp)$ with the replacements $u_\parallel=u\mu$ and
$u_\perp=u\sqrt{1-\mu^2}$ and retaining only the pitch angle scattering part in 
(\ref{eq:dFdt5}), one obtains 
\begin{eqnarray}
{d F(u,\mu)\over
dt}&=&{\partial\over\partial\mu}\Biggl[(1-\mu^2)D{\partial\over\partial\mu}\Biggr]F(u,\mu),\nonumber\\
D&=&{\pi\over4}{\Omega\over\gamma}{\delta B^2_r\over B^2},
\end{eqnarray}
where 
\begin{equation}
{\delta B^2_r\over\mu_0}= 
{k_r\over V}\int {dk_\perp k_\perp\over (2\pi)^2}
{\delta B^2_k\over\mu_0}\Biggr|_{k_r}
\end{equation}
is the density of magnetic energy at the resonant wavenumber $k_r=\Omega/cu|\mu|$.
Using (\ref{eq:fcr}), it can be shown that the resonant diffusion does not contribute to
change in $\langle u^2_\parallel\rangle$ and $\langle u^2_\perp\rangle$. Thus, 
the rate of change in parallel and perpendicular pressures is determined 
by nonresonant diffusion. 

The rate of change in the streaming speed is 
\begin{equation}
{dv_{_{\rm CR}}\over dt}=-\bar{D} v_{_{\rm CR}},
\label{eq:vcr2}
\end{equation}
where
\begin{equation}
\bar{D}={1\over2}\int^1_{-1}d\mu (1-\mu^2)D
\approx {\pi^2\over4}{\Omega\over\gamma}{\delta B^2_r\over B^2}{1\over k_0r_{g0}},
\label{eq:Db}
\end{equation}
where $\delta B_r$ is set the value at $\mu\sim \pi/2$.
The approximation in (\ref{eq:Db}) is obtained by writing 
$\delta B^2_k=(2\pi)^3V\delta B^2_r\delta(k_\perp)\delta(k_r-k_0)/k_\perp$, i.e. one assumes
that the magnetic fluctuations are centered at the wave number $k_0$. 
Here $r_{g0}$ is the gyroradius of CRs with an average Lorentz factor $\gamma_0=\langle\gamma\rangle$.

Since the rate of the change in streaming motion due to resonant diffusion is $
(dv_{_{\rm CR}}/dt)_r\propto \Omega/\gamma$ and  the corresponding rate due 
to nonresonant diffusion is $(dv_{_{\rm CR}}/dt)_{nr}\propto \Gamma/\gamma^2$, one has
\begin{equation}
\left|{(dv_{_{\rm CR}}/dt)_r\over
(dv_{_{\rm CR}}/dt)_{nr}}\right|\sim5\times 
10^{-2}{\gamma\Omega\over\Gamma}\left({\delta B^2_r\over\delta B^2}\right)
\left({B^2_*\over B^2}\right){1\over k_0r_{g0}}.
\label{eq:ratio}
\end{equation}
Here $\delta B$ is the typical amplitude of the magnetic fluctuations generated 
by the nonresonant instability. The right-hand side is larger than 1 for 
$k_0 r_{g0}>20(\Gamma/\gamma_0\Omega)$, where
one assumes $\delta B_r\sim \delta B$ and $B_*/B \sim k_0r_{g0}$. This condition is satisfied 
for the typical parameters relevant for SN shocks (cf. Sec 5).
If one writes $\delta B\propto\exp(\Gamma t)$ and takes the limit $t\gg1/\Gamma$, 
Eq (\ref{eq:vcr2}) can be solved to yield
\begin{equation}
{\delta B^2\over B^2}\approx {12\over\pi}{\Gamma\gamma\over\Omega}k_0r_{g0}\ln\left({v_{_{\rm CR}0}
\over v_{_{\rm CR}}}\right).
\label{eq:sat}
\end{equation}

\section{Magnetic fields in SNRs}

There are a number of SNRs whose post-shock magnetic fields have been estimated 
from X-ray observations and found to be in the range $\geq 10^{-8}\,\rm T$, 
much higher than the typical value in the ISM ($\sim 10^{-10}\,\rm T$)
\citep{vl03,vetal05}. Although one cannot rule out the possibility that such 
strong field is due to compression of strong pre-existing magnetic fields in the stellar 
wind of the progenitor star, the observations favor the interpretation that the 
magnetic fields are due to an instability at the shock. Observations show that the 
magnetic field strength strongly depends on the shock speed \citep{vl03,vetal05}. 
The post-shock magnetic energy density in SNRs whose shock speeds have been estimated show 
a tendency $\delta B^2\propto v^\delta_s$, where $v_s$ is the shock speed \citep{vetal05,v08}. 
The argument that the magnetic pressure is proportional to the ram pressure ($\propto v^2_s$)
at the shock front suggests $\delta=2$, while \citet{b04}'s model predicts 
$\delta=3$. So far, current observations cannot differentiate these two because of 
the narrow dynamic range of the observed properties of SN shocks \citep{v08}.

\subsection{SN shocks}

We consider a forward shock traveling at a velocity $v_s$. The shock undergoes
initially a free expansion phase and after sweeping up sufficient mass, it 
enters a Sedov phase. In the Sedov
phase, the velocity is given by $v_s=2R_s/5t$, where $R_s$ is the shock radius.
The total kinetic energy of the shock is $E=(4/3)R^3_s\rho_0 v^2_s$,
where $\rho_0$ is the density of the ISM. On eliminating $R_s$ in favour of $t$, 
the shock velocity becomes
\begin{equation}
v_s=\left({6E\over125\rho_0}\right)^{1/5}t^{-3/5}.
\end{equation}
One can show that development of the nonresonant instability is faster
in the early Sedov phase. The maximum growth rate is $\propto J^{1/2}_{CR}$, 
which can be written in terms of the ram pressure in the up-stream region.
Thus, one expects the growth rate to be a strong function 
of the shock speed. Since the ram pressure is $\sim \rho_0 v^2_s$,
the CR pressure can be expressed as $P_{CR}=\eta_s\rho_0 v^2_s$, where $\eta_s\leq1$ is the
acceleration efficiency. We consider precursor CRs with the initial streaming velocity
$v_{_{\rm CR0}}\sim v_s$. Since $P_{CR}=n_{CR}mc\langle uv\rangle/3$, one has
$J_{CR}=3\eta_s q\rho_0v^2_sv_{CR}/(mc\langle uv\rangle)
\approx 3\eta_s q\rho_0v^2_sv_{CR}/[mcu_1\ln(u_2/u_1)]$, where $\langle uv\rangle
\approx cu_0$ and $u_0$ is the average momentum given by 
\begin{equation}
u_0=\int^\infty_0ug(u)du=u_1{p-1\over p-2}\,{1-(u_2/u_1)^{2-p}\over
1-(u_2/u_1)^{1-p}}.
\label{eq:p0}
\end{equation}
The average Lorentz factor of CRs is $\gamma_0=(1+u^2_0)^{1/2}$.
For $p=2$ and $u_2\gg u_1$, one has $u_0=u_1\ln(u_2/u_1)$.
Therefore, the faster the shock speed the higher the growth rate.

\subsection{Saturated magnetic field vs shock speed}

One may estimate the maximum magnetic field, $\delta B_m$, of the wave by equating 
the magnetic pressure to the CR pressure, $\delta B^2_m/2\mu_0=P_{CR}$. This leads
to $\delta B_m\approx (2\mu_0\eta_s\rho_0)^{1/2} v_s$, or
\begin{equation}
\delta B_m \approx
3\times10^{-7}\eta^{1/2}_s\left({n_0\over10^6\,{\rm
m}^{-3}}\right)^{1/2}\left({v_s\over5\times10^6\,{\rm
m}\,{\rm s}^{-1}}\right)\,{\rm T}. 
\label{eq:dBm}
\end{equation}
Eq (\ref{eq:dBm}) is a rather optimistic estimate and calculation of the saturation 
due to the feedback effects of the instability on the streaming anisotropy
generally leads to a lower saturation level than (\ref{eq:dBm}).

One can calculate the saturated magnetic field due to the diminishing of the streaming anisotropy
from (\ref{eq:sat}). This involves calculation of the cut-off streaming speed, 
denoted by $v'_{_{\rm CR}}$, at which the instability turns off. From (\ref{eq:condition}), one obtains
\begin{equation}
v'_{_{\rm CR}}\approx {5c\over24\eta_s}{U_B\over\rho_0v^2_s}k_\parallel r_{g0},
\label{eq:vcrf}
\end{equation}
where $U_{B}=B^2/2\mu_0$ is the energy density of the mean magnetic field.
At a given parallel wavenumber $k_\parallel$, wave growth stops when the streaming 
speed is reduced to $v_{_{CR}}\leq v'_{_{\rm CR}}$. One may assume $v_{_{CR0}}\sim v_s$.
Eq (\ref{eq:vcrf}) gives
\begin{eqnarray}
{v_{_{\rm CR0}}\over v'_{_{\rm CR}}}&\approx&
5\times10^5 {\eta_s\over k_\parallel r_{g0}}
\left({v_s\over5\times10^6\,{\rm m}\,{\rm s}^{-1}}\right)^3\left({10^{-10}\,{\rm T}\over B}\right)^2
\nonumber\\
&&\times
\left({n_0\over 10^6\,{\rm m}^{-3}}\right).
\label{eq:vratio}
\end{eqnarray}
For convenience, one assumes here that CRs and ions in the ISM are all protons.

The growth rate (\ref{eq:Gam2}) is estimated to be 
$\Gamma\approx (6\eta_sk_0r_{g0})^{1/2}(v_s/c)^{3/2}\Omega/\gamma_0$.
From (\ref{eq:sat}) one obtains
\begin{equation}
{\delta B^2\over B^2}\approx 
{12\over\pi}\left(6\eta_s\right)^{1/2}
\left({v_s\over c}\right)^{3/2}\left(k_0r_{g0}\right)^{3/2}\ln\left({v_{_{\rm CR}0}\over v'_{_{\rm CR}}}\right).
\label{eq:sat2}
\end{equation}
Eq (\ref{eq:sat2}) implies that significant amplification $\delta B^2/B^2\gg1$ 
is possible if one assumes a large $k_0r_{g0}$. However, one should
point out that the quasilinear theory is based on the weak turbulence assumption
that requires the relevant wave fluctuctuations be small. Nonetheless, since the saturation
mechanism discussed is quite generic, i.e. the saturation is attributed to the 
reduction in CR streaming, one expects that extropolation of Eq (\ref{eq:sat2}) to
the $\delta B^2/B^2\gg1$ regime would still provide a reasonable estimate for the 
saturation. To estimate the upper limit to the saturation one assumes that the 
Larmor radius is limited by the scale length of the 
acceleration region, which is assumed to be the shock's radius, $R_s$. The
characteristic wavenumber $k_0$ is limited by the condition 
(\ref{eq:condition}), which gives $k_0\sim (\omega_{p,{\rm CR}}/c)(v_{_{\rm CR}}/v_A)$,
where $\omega_{p,{\rm CR}}$ is the plasma frequency of CRs.
This corresponds to $(k_0r_{g0})_{\rm max}\sim 10^2-10^3$.
For $\eta_s=0.1$, $v_s=5\times10^6\,{\rm m}\,{\rm s}^{-1}$ and $k_0r_{g0}=10^2$, one has
$\delta B^2/B^2\sim 39$. The saturated magnetic field can exceed the 
background field by a modest factor. 
If the CR pressure is a fixed fraction of the ram pressure, one has 
$n_{_{\rm CR}}\gamma_0\propto v^2_s$. Since the shock radius can be
expressed as $R_s=(3E/4\rho_0v^2_s)^{1/3}$, one has 
$k_0R_s\propto v^{1/3}_s$, where $k_0\propto v_{_{\rm CR}}
\sim v_s$. Eq (\ref{eq:sat2}) implies that the energy density 
of the saturated magnetic field 
increases with the shock speed as $\delta B^2\propto v^2_s$.
Remarkably, this predicts the same shock-speed scaling as the one from 
the equipartition argument, i.e. the magnetic pressure 
equal to the CR pressure (cf. Eq \ref{eq:dBm}). 

\section{Conclusions and discussion}

We develop a kinetic version of the CR streaming instability of
\citet{b04}, who treated it using MHD. We show that the nonresonant growth
is not due to the CRs themselves, but rather to a current in the background plasma
that neutralizes the current due to streaming CRs. Saturation of the CR streaming 
instability is discussed in the quasilinear theory. Growth of the nonresonant instability
leads to particle diffusion that suppresses the instability. In the quasilinear diffusion formalism,
the saturated magnetic field can be calculated analytically.  We consider
both resonant and nonresonant diffusion, with the latter being
treated in a similar way to that used for the firehose instability \citep{ss64,d72}. 
In our model the CR pressure is set to a fixed fraction of the ram pressure 
at the shock front. It is shown that evolution of the 
parallel and perpendicular pressures is due to nonresonant diffusion. The direction
of the evolution is that the parallel energy increases and 
perpendicular energy decreases. This is in contrast to the firehose instability which causes
the perpendicular pressure to increase and parallel pressure to decrease.
One should note the difference in the sources of free energy that drive the
instabilities; the free energy for the firehose instability is an excess of
parallel pressure while the free energy for the nonresonant instability considered here is 
streaming motion of CRs. It is shown that saturation is determined 
by resonant diffusion which reduces the streaming motion.
The saturated magnetic fluctuations can exceed the background field provided 
that the shocks are sufficiently fast and saturation occurs at short wavelength
$1/k_0\ll1$.

There have been numerical simulations of the streaming instability in both the MHD 
approximation and kinetic theory, which confirm rapid growth of the instability
under the physical conditions of young SN shocks. However, these simulations predict different
saturation levels. One of the main difficulties in numerical simulation of the instability is the
huge difference between the CR number density and background plasma density, with
$n_{_{\rm CR}}/n_0\sim 10^{-5}$. A particular approximation is usually adopted to make numerical
simulation of such system feasible. For example, in \citet{netal08}'s numerical model, 
the density ratio is set artificially to a large value. Because the existing numerical models 
use different approximations, and it is unrealistic to make
meaninful comparison among their results. 
MHD simulations generally predict a saturation at strong 
magnetic fields \citep{b04,zetal08}, while kinetic particle-in-cell (PIC) 
simulations appear to predict much lower saturated magnetic fields \citep{netal08,rs08}.  
The analytical calculation presented here predicts a relatively modest amplification 
that gives rise to magnetic field fluctuations similar to the level predicted by the 
recent kinetic simulations \citep{rs08}.



\appendix

\section{Quasilinear diffusion}

Using the similar method discussed in \citet{d72},
we write the distribution function as $f\approx f^{(0)}+f^{(1)}$ with $f^{(1)}$ a linear
function of electric and magnetic fluctuations, $\delta\bE$ and $\delta\bB$,
of the wave. A formal approach to determine the feedback effect of the wave on the 
particle distribution involves calculation of time-dependence of the distribution averaged
over the random phase, denoted by $F\equiv \langle f^{(0)}\rangle_R$, where 
$\langle\cdots\rangle_R$ is the random phase average.

\subsection{General formalism}

From the Vlasov equation one obtains
\begin{equation}
{d F\over dt} +q\left\langle(\delta \bE+\bv\times\delta\bB)\cdot{\partial f^{(1)}\over\partial \bp}
\right\rangle_R=0,
\label{eq:ve}
\end{equation}
where $d/dt=\partial/\partial t+\bv_D\cdot\partial/\partial\bx$, $\bv_D$ is the drift velocity.
Let $\xi\equiv(\delta \bE, \delta\bB, f^{(1)})$ and the corresponding
spatial Fourier transform be $\xi_k\equiv(\delta \bE_k, \delta\bB_k, f^{(1)}_k)$, defined by
\begin{equation}
\xi(\bx,t)=\int{d\bk\over(2\pi)^3}\xi_k(t)\exp\Bigl(i\bk\cdot\bx\Bigr),
\end{equation}
where we assume WKB for the time dependence
\begin{equation}
\xi_k(t)\propto \exp\Bigl[-i\psi_k(t)\Bigr],\quad
\psi_k(t)=\int^t_0\Bigl[\omega_k(t)+i\Gamma(t)\Bigr]dt'.
\end{equation}
Here $\omega_k$ is the wave frequency and $\Gamma$ is the growth rate.
One adopts the convention $\omega_{-k}=-\omega_k$, $\xi_{-k}=\xi^*_k$.

In cylindrical coordinates, the velocity can be expressed as
$\bv=(v_\perp\cos\phi,v_\perp\sin\phi,v_\parallel)$, where $\phi$ is the gyrophase.
The 3-axis is assumed to be along the magnetic field. For a transverse wave propagating 
parallel to the magnetic field, one has $\delta E_{\pm}\equiv
\delta E_{kx}\pm \delta E_{ky}/i=\mp (i\omega/k_\parallel)
(\delta B_{kx}\pm \delta B_{ky}/i)\equiv \mp (i\omega/k_\parallel)\delta B_\pm$.
In the limit $k_\perp v_\perp/\Omega\ll1$, the perturbation of the distribution
can be written as
\begin{equation}
f^{(1)}_k={\textstyle{1\over2}}\biggl( f_+e^{i\phi}+f_-e^{-i\phi}\biggr).
\label{eq:f1}
\end{equation}
From the linearized Vlasov equation one obtains
\begin{eqnarray}
f_\pm&=&{\pm q\delta B_\pm\over
\omega+i\Gamma-k_\parallel v_\parallel\pm\Omega/\gamma}\nonumber\\
&&\times\Biggl[v_\perp{\partial\over\partial p_\parallel}-
\biggl(v_\parallel-{\omega+i\Gamma\over k_\parallel}\!\biggr){\partial\over\partial p_\perp}\!\Biggr]F.
\label{eq:fpm}
\end{eqnarray}
Substituting (\ref{eq:f1}) and (\ref{eq:fpm}) into (\ref{eq:ve}), one derives
\begin{eqnarray}
{dF\over dt}&=&-{q\over4iV}\int{d\bk\over(2\pi)^3}\Biggl[v_\perp{\partial\over\partial p_\parallel}
\nonumber\\
&&-
\biggl(v_\parallel-{\omega-i\Gamma\over k_\parallel}\biggr){1\over p_\perp}{\partial\over\partial p_\perp}p_\perp
\Biggr]\nonumber\\
&&\times\Bigl(\delta B^*_+f_+-\delta B^*_-f_-\Bigr),
\end{eqnarray}
where one uses
\begin{equation}
\langle\delta\bB_k\cdot\delta\bB_{k'}\rangle_R={(2\pi)^3\over V}|\delta B_k|^2\delta(\bk+\bk').
\end{equation}
With $|\delta B_k|^2=|\delta B_+|^2=|\delta B_-|^2$, one obtains
\begin{eqnarray}
{dF\over dt}&=&-{q\over2iV}\int{d\bk\over(2\pi)^3}|\delta B_k|^2\nonumber\\
&&\times\Biggl[v_\perp{\partial\over\partial p_\parallel}-
\biggl(v_\parallel-{\omega-i\Gamma\over k_\parallel}\biggr){1\over p_\perp}{\partial\over\partial p_\perp}p_\perp
\Biggr]\nonumber\\
&&\times
{\omega+i\Gamma-k_\parallel v_\parallel\over(\omega+i\Gamma-k_\parallel v_\parallel)^2-\Omega^2/\gamma^2}\nonumber\\
&&\times
\Biggl[v_\perp{\partial\over\partial p_\parallel}-
\biggl(v_\parallel-{\omega+i\Gamma\over k_\parallel}\biggr){\partial\over\partial p_\perp}
\Biggr]F.
\label{eq:dFdt3}
\end{eqnarray}
For practical purposes, we set $\omega\sim 0$ here.

\subsection{Nonresonant diffusion}

The quasilinear diffusion equation in the long wavelength approximation can be obtained 
using the following expansion in $1/\Omega$:
\begin{equation}
{\omega+i\Gamma-k_\parallel v_\parallel\over(\omega+i\Gamma-k_\parallel
v_\parallel)^2-\Omega^2/\gamma^2}
\approx {k_\parallel v_\parallel\over\tilde{\Omega}^2}-{i\Gamma\over\tilde{\Omega}^2},
 \label{eq:exp1}
\end{equation}
with $\tilde{\Omega}=\Omega/\gamma$.
We consider a more general expansion in $\Gamma/|k_\parallel v_\parallel-\tilde{\Omega}|\gg1$:
\begin{eqnarray}
{i\Gamma-k_\parallel v_\parallel\over(i\Gamma-k_\parallel v_\parallel)^2-\tilde{\Omega}^2}
&\approx&{1\over k^2_\parallel v^2_\parallel-\tilde{\Omega}^2}\nonumber\\
&&\times\Biggl(
i\Gamma-k_\parallel v_\parallel-{2i\Gamma k^2_\parallel v^2_\parallel\over k^2_\parallel v^2_\parallel-\tilde{\Omega}^2}\Biggr).
\end{eqnarray}
Integration on the right-hand side of (\ref{eq:dFdt3}) includes terms $\propto k_\parallel$ or $\propto 1/k_\parallel$.
For turbulent spectra with symmetry $k_\parallel\to-k_\parallel$, one has
\begin{equation}
\int^\infty_{-\infty}dk_\parallel\,\left({\rm terms}\propto k_\parallel\,{\rm or}\,\propto 1/k_\parallel\right)=0,
\end{equation}
and only terms $\propto i\Gamma$ remain.

\subsection{Resonant diffusion}

The standard procedure to write down the resonant diffusion involves setting $\Gamma=0$.
Eq (\ref{eq:dFdt3}) can be expressed in the following familiar form
\begin{eqnarray}
{dF\over dt}&=&{\partial\over\partial p_\parallel}\Biggl(
D_{\parallel\parallel}{\partial\over\partial p_\parallel}+D_{\parallel\perp}{\partial\over\partial p_\perp}
\Biggr)\nonumber\\
&&+{1\over p_\perp}{\partial\over\partial p_\perp}\Biggl(
D_{\perp\parallel}{\partial\over\partial p_\parallel}+D_{\perp\perp}{\partial\over\partial p_\perp}\Biggr),
\label{eq:dFdt5}
\end{eqnarray}
where $D_{\alpha\beta}$ are the diffusion coefficients.


\begin{thebibliography}{22}
\bibitem[\protect\citeauthoryear{Achterberg}{1983}]{a83}
Achterberg, A., 1983, A\&A, 119, 274 
\bibitem[\protect\citeauthoryear{Amato \& Blasi}{2008}]{ab08}
Amato, E., Blasi, P., 2008, MNRAS, in press
\bibitem[\protect\citeauthoryear{Bell}{1978}]{b78}
Bell, A. R., 1978, MNRAS, 182, 147
\bibitem[\protect\citeauthoryear{Bell}{2004}]{b04}
Bell, A. R., 2004, MNRAS, 353, 550
\bibitem[\protect\citeauthoryear{Bell \& Lucek}{2001}]{bl01}
Bell, A. R., Lucek, S. G., 2001, MNRAS, 321, 433 
\bibitem[\protect\citeauthoryear{Davidson}{1972}]{d72}
Davidson, R. C., 1972, Methods in Nonlinear Plasma Theory (Academic Press: New York)
\bibitem[\protect\citeauthoryear{Drury}{1983}]{d83}
Drury, L. O'C., 1983, Rep. Prog. Phys., 46, 973
\bibitem[\protect\citeauthoryear{Eichler}{1984}]{e84}
Eichler, D., 1984, ApJ, 277, 429
\bibitem[\protect\citeauthoryear{Hillas}{2006}]{h06}
Hillas, A. M., 2006, J. Phys. Conf. Series 47, 168
\bibitem[\protect\citeauthoryear{Kulsrud \& Pearce}{1969}]{kp69}
Kulsrud, R., Pearce, W. P., 1969, ApJ, 156, 445
\bibitem[\protect\citeauthoryear{Lagage \& Cesarsky}{1983a}]{lc83a}
Lagage, P. O., Cesarsky, C. J., 1983a, A\&A, 118, 223
\bibitem[\protect\citeauthoryear{Lagage \& Cesarsky}{1983b}]{lc83b}
Lagage, P. O., Cesarsky, C. J., 1983b, A\&A, 125, 249
\bibitem[\protect\citeauthoryear{Machabeli et al.}{2000}]{metal00}
Machabeli, G. Z., Luo, Q., Melrose, D. B., Vladmirov, S., 2000, MNRAS, 312, 51
\bibitem[\protect\citeauthoryear{McKenzie \& V\"olk}{1982}]{mv82}
McKenzie, J. F., V\"olk, H. J., 1982, A\&A, 116, 191
\bibitem[\protect\citeauthoryear{Melrose \& Wentzel}{1970}]{mw70}
Melrose, D. B., Wentzel, D. G., 1970, ApJ, 457, 476
\bibitem[\protect\citeauthoryear{Melrose}{2005}]{m05}
Melrose, D. B., 2005, AIP Proceedings, 781, 135
\bibitem[\protect\citeauthoryear{Melrose}{1986}]{m86}
Melrose, D. B., 1986, Instabilities in Space and Laboratory Plasmas, Cambridge Press 
\bibitem[\protect\citeauthoryear{Niemiec, Pohl, Stroman, \& Nishikawa}{2008}]{netal08}
Niemiec, J., Pohl, M., Stroman, T., Nishikawa, K. I., 2008, ApJ, 684, 1174
\bibitem[\protect\citeauthoryear{Reville, Kirk \& Duffy}{2006}]{retal06}
Reville, B., Kirk, J. G., Duffy, P., 2006, Plasma Phys. Control. Fusion, 48, 1741
\bibitem[\protect\citeauthoryear{Riquelme \& Spitkovsky}{2008}]{rs08}
Riquelme, M. A., Spitkovsky, A., 2008, ApJ, submitted
\bibitem[\protect\citeauthoryear{Shapiro \& Shevchenko}{1964}]{ss64}
Shapiro, V. D., Shevchenko, V. I., 1964, Sov. Phys. JETP 18, 1109
\bibitem[\protect\citeauthoryear{Skilling}{1975}]{s75}
Skilling, J., 1975, MNRAS, 172, 557
\bibitem[\protect\citeauthoryear{Tanaka, et al.}{2008}]{tetal08}
Tanaka, T., et al., 2008, ApJ, in press
\bibitem[\protect\citeauthoryear{Vink}{2008}]{v08}
Vink, J., 2008, astro-ph 0810.3680
\bibitem[\protect\citeauthoryear{Vink \& Laming}{2003}]{vl03}
Vink, J., Laming, J. M., 2003, ApJ, 584, 758
\bibitem[\protect\citeauthoryear{V\"olk, Berezhko \& Ksenofontov}{2005}]{vetal05}
V\"olk, H., Berezhko, E. G., Ksenofontov, L. T., 2005, A\&A, 433, 229
\bibitem[\protect\citeauthoryear{V\"olk, Ksenofontov \& Berezhko}{2008}]{vetal08}
V\"olk, H., Ksenofontov, L. T., Berezhko, E. G., 2008, astro-ph0809.2432
\bibitem[\protect\citeauthoryear{V\"olk \& McKenzie}{1981}]{vm82}
McKenzie, J. F., V\"olk, H., 1982, 17th Int. Cosmic Ray Conf., Vol. 9, 242
\bibitem[\protect\citeauthoryear{Weibel}{1959}]{w59}
Weibel, E. S., Phys. Rev. Lett. 2, 83
\bibitem[\protect\citeauthoryear{Zirakashvili, Ptuskin \& V\"olk}{2008}]{zetal08}
Zirakashvili, V. N., Ptuskin, V. S., V\"olk, H. J., 2008, ApJ, 678, 255
\end{thebibliography}
\end{document}